\documentclass[runningheads]{llncs}
\usepackage[T1]{fontenc}
\usepackage{todonotes}
\usepackage{booktabs}
\usepackage{url}
\usepackage{array}
\usepackage{changepage} 

\newcommand{\benefit}[0]{\item[$+$]}
\newcommand{\liability}[0]{\item[$-$]}

\usepackage{graphicx}
\begin{document}

\title{Patterns for Teaching Agile with Student Projects -- Team and Project Setup}

%
%\titlerunning{Abbreviated paper title}
% If the paper title is too long for the running head, you can set
% an abbreviated paper title here
%
\author{Daniel Pinho\inst{1,2} \and
Petr Pícha\inst{3} \and\\
Filipe F. Correia\inst{1,2} \and
Přemek Brada\inst{3}
}
\authorrunning{D. Pinho et al.}
% First names are abbreviated in the running head.
% If there are more than two authors, 'et al.' is used.
%

\institute{Faculty of Engineering, University of Porto, Porto, Portugal \and
INESC TEC, Porto, Portugal \\
\email{\{daniel.pinho,filipe.correia\}@fe.up.pt}\\
\and
University of West Bohemia in Pilsen, Pilsen, Czechia
\\\email{\{ppicha,brada\}@kiv.zcu.cz}
}

\maketitle              % typeset the header of the contribution
%
%\linenumbers            % starts counting line numbers (FOR DRAFT VERSIONS ONLY)

\begin{abstract}
Higher education courses teaching about agile software development (ASD) have increased in commonality as the ideas behind the Agile Manifesto became more commonplace in the industry. However, a lot of the literature on how ASD is applied in the classroom does not provide much actionable advice, focusing on frameworks or even moving beyond the software development area into teaching in an agile way. We, therefore, showcase early work on a pattern language that focuses on teaching ASD practices to university students, which stems from our own experiences as educators in higher education contexts. We present five patterns, specifically focused on team and project setup phase: \textsc{Capping Team Size}, \textsc{Smaller Project Scope}, \textsc{Business Non-Critical Project}, \textsc{Self-assembling Teams}, and \textsc{Team Chooses Topic} as a starting point for developing the overall pattern language.

\end{abstract}

%------------------------
%  BEGIN DOCUMENT 
%  (template text, containing information on tables, figures, acknowledgements and conflict of interests has been moved to template-referece)
%------------------------

\section{Introduction}\label{sec:introduction}

As the principles behind the Agile Manifesto~\cite{beckManifestoAgileSoftware2001} grew in popularity in the software industry, higher education courses on agile software development~(ASD) have become more and more common~\cite{rico2009UseAgileMethods}. A wide variety of literature on how Agile is found in the classroom does not provide directly actionable advice, focusing on frameworks~\cite{tamayoavila2022ImprovingTeamworkAgile}, exploring team dynamics~\cite{saeter2024RoleTeamComposition}, and even moving beyond the software development area~\cite{lopez-alcarria2019SystematicReviewUse} into teaching in an agile way.

In this paper, we present the first steps towards a pattern language focused on courses teaching ASD practices to university students, showcasing a subset of patterns in the language. These patterns come from our own experiences as educators at the University of Porto and the University of West Bohemia in Pilsen, working with students in bachelor's and master's software engineering courses.

Our working definition of "agile courses" (and similar) within the context of this and following papers on the pattern language covers all higher education software engineering courses that include teaching the principles and/or practices of agile development, from general adaptability of the process to change to specifics, like sprint retrospectives, stand-ups, backlog refinement, and so on. We see learning Agile in this sense as a valuable outcome for the students, not only because agile methodologies are prevalent in practice, but also as the ability to adopt one's process to fit current circumstances while keeping on the path to set goals is a good skill to have in general. As such, proper guidance for teachers on how to facilitate this learning is a worthwhile effort and sorely needed.

The patterns in this work target an audience of university professors and teachers who either teach project-based software engineering courses with agile elements (or a whole process) looking to improve or validate their approach, or are building a similar course looking for guidance and support. We chose this target audience for the patterns because they, the educators, are the ones who have the most control over the students' working environment and, as such, can take action directly in an impactful way.

However, the specific patterns described in this paper are not strictly Agile-related and can be used in any course with collaborative student projects. This is due to these patterns dealing with the stage of setting up the teams, gathering project topics and assigning them. Agile practices are not in strong effect in this phase but it is a necessary ground work for further patterns in our proposed language, which will be the focus of future work. 

The remainder of this work is structured as follows. 
Section~\ref{sec:related-work} explores related work and background for this paper. Section~\ref{sec:pl-overview} introduces the pattern language, describes its sources~\ref{ssec:courses}, the template used~\ref{ssec:template} and the common context for all the patterns within~\ref{ssec:common-context}. Section~\ref{sec:patterns} then showcases the five patterns we are presenting: \textsc{Capping Team Size}~(\ref{pattern:team-size}) sets a boundary on the student project team. \textsc{Smaller Project Scope}~(\ref{pattern:smaller-scope}) discusses how large in scope should projects be to maintain reachable learning objectives. \textsc{Business Non-Critical Project}~(\ref{pattern:business-non-critical-project}) advises customers (internal and/or external) to propose project topics that are not business-critical but still bring value. \textsc{Self-Assembling Teams}~(\ref{pattern:self-assembling}) argues for giving students the freedom to form their own teams. \textsc{Team Chooses Topic}~(\ref{pattern:team-chooses-topic}) advocates that students have a say in the topics they work on to increase their buy-in.
Finally, in Section~\ref{sec:conclusions} we draw our conclusions from this work and propose future research.
\section{Related work}\label{sec:related-work}

We found a lack of existing literature on teaching agile practices and processes with projects that would present guidance in an actionable manner, in pattern form or otherwise. The following presents some examples of the closest related work we found and how it differs from our goals.

As a basis, Sutherland and Coplien~\cite{sutherland2019scrum} gives a good overview of what we want the students to learn in general, but is not specific to teaching via student projects.

Jacobs~\cite{jacobs2023supporting} has done work on patterns for teachers to support software engineering students in their projects. Though it does includes some advice on agile practices (like retrospectives, daily stand-ups and product delivery meetings with the customer), Agile is not its central focus.

They further focused on specifically on the category of patterns for grading student team projects~\cite{jacobs2024grading}. Their results are both more narrow, on the category level, and more general, in the sense of not focusing on Agile, than ours.

K{\"o}ppe~\cite{koppe2012learning}, and subsequently de Cortie et al.~\cite{de2013learning} have published a two-part work on learning patterns for group assignments. Though very valuable and providing overall guidance, the patterns are written from the students' perspective and do not focus on agile projects.

Tamayo Avila et al.~\cite{tamayoavila2022ImprovingTeamworkAgile} performed literature reviews and a correlational study to propose an enhancement for
a pre-existing framework for fostering student learning and performance with agile practices. It is, however, more narrowly focused on improving internal team cohesion, rather than the overall guide on teaching through student projects that we aim for.

Sæter et al.~\cite{saeter2024RoleTeamComposition} performed a case study to explore team dynamics in agile student teams and how gender composition affects teamwork.
We were not able to find a more general study of similar kind absent the gender aspect.

López-Alcarria et al~\cite{lopez-alcarria2019SystematicReviewUse} performed a systematic literature review to analyse how agile methods can support education for sustainable development. They found that agile principles and methods overlap with education for sustainable development competencies, and note that educators can establish parallels between agile contexts in the industry and the classroom. However, they do not provide substantially detailed and actionable guidance.

Other sources focus on applying agile methodologies to teaching and course design activities themselves~\cite{fernanda2018agile}, provide experience reports and case studies on teaching Agile~\cite{devedvzic2010teaching}, or teaching agile principles and practices via lab courses~\cite{schroeder2012teaching} or other means~\cite{kropp2014teaching}. None of which is directly applicable or conducive to our goals.

Our work in this and subsequent papers is part of a wider efforts with other authors and the community at large to catalogue, evolve and maintain patterns related to teaching software engineering in higher education courses with student projects. The results of this work are kept in an online, publicly accessible, text-based repository\footnote{\url{https://github.com/ReliSA/STePSEnHECs-PaCt}} and open to contributions.

\section{Pattern language overview}\label{sec:pl-overview}

The entire pattern language is focused specifically on agile software project execution practices in the context of student team projects in higher education. In other words, it strives to provide comprehensive guidance for teachers on how to set up and run collaborative student project courses to simultaneously provide the opportunity for students for a hands-on experience with agile practices whilst also keeping the projects manageable from both the teams' and staff perspective.

As a result, not all patterns in the language are strictly agile in nature or tie back to its principles directly. To serve as a blueprint for courses, the language necessarily covers even methodology-agnostic activities, and connective tissue and grounding aspects. This is especially true for the specific patterns described in this paper, dealing with team and project setup. The agile practices the overall language does cover are the ones that need to be tailored to our context from typical form in industry practice. We do not intend to cover aspects of agile projects that can be applied equally in educational environment as in practice, as there is ample material on general agile guidance already in existence.

The patterns were mined from our experiences as teachers and educators involved in software engineering higher education courses, with complementary information gathered from the literature. We discussed the processes we employ when preparing and teaching our courses and documented the practices where we identified similarities between our approaches. We also confirmed the wider-spread usage of the patterns by surveying some of our colleagues running similar courses in their institutions. As the starting point were specifically the courses we teach ourselves (see~\ref{ssec:courses}), alternative solutions and even different permutations of the whole language are almost certain to exist. However, we cannot fully confirm them as patterns with enough observations and thus do not include them in our language at this time.

Although we were able to mine twenty distinct patterns so far, we present only five in fully described form in this paper. The full set can be found captured in patlet form in Appendix~\ref{ap:patlets}, organized in several categories, and will be explored further in future work. We did this to describe our gathered knowledge more thoroughly and gradually.

\begin{figure}[htbp]
\centering
\includegraphics[width=\textwidth]{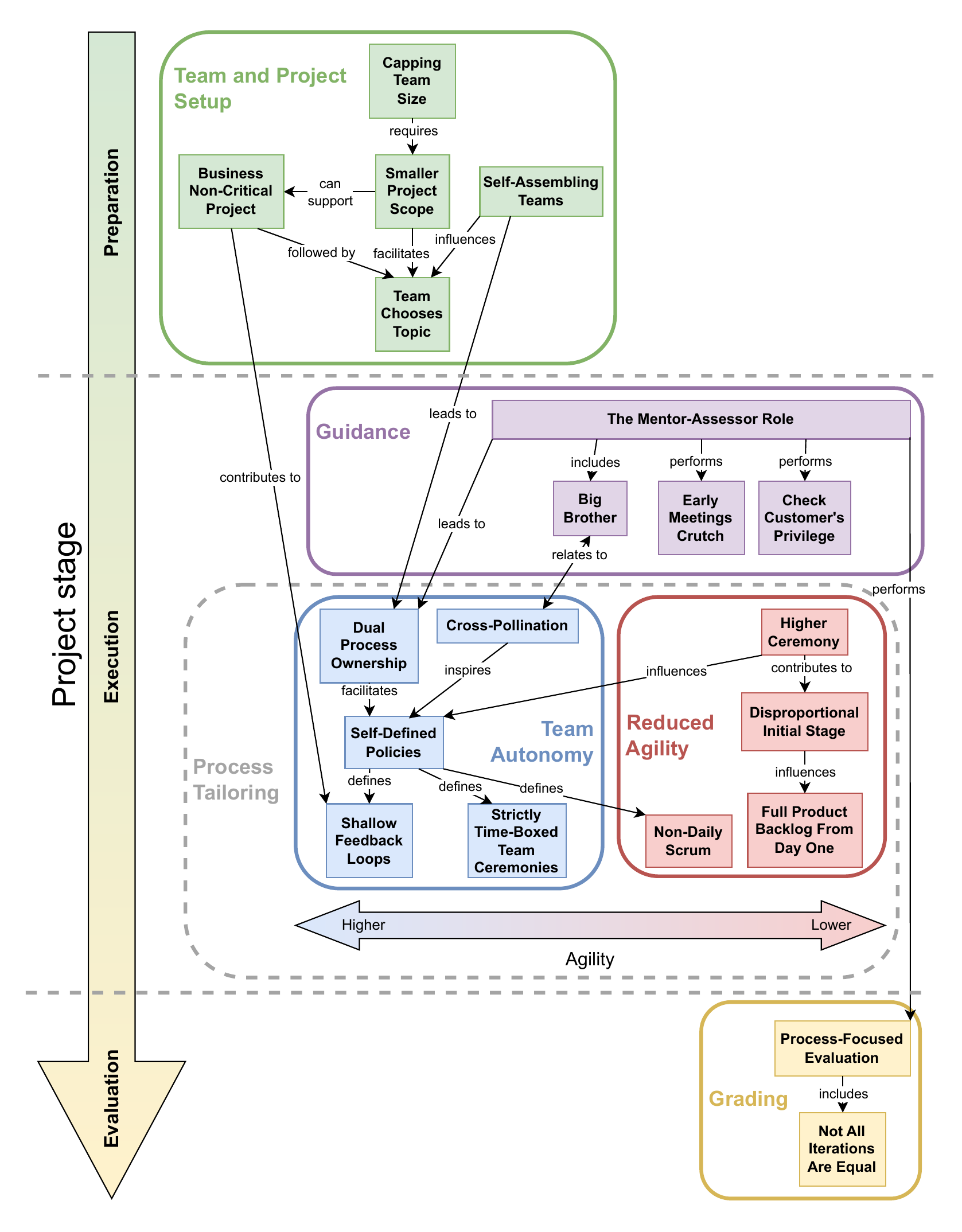}
\caption{Current state of the pattern language.} 
\label{fig-pat-lan}
\end{figure}

Figure~\ref{fig-pat-lan} shows the entire current state of our pattern language, organized roughly into categories on the top level by \textit{Project stage}. This paper focuses on the \textit{Team and Project Setup} (i.e., \textit{Preparation}) stage, which ends by pairing teams with assigned topics. In the \textit{Execution} stage, the patterns deal mainly with \textit{Process Tailoring} (super-category) overseen by \textit{Guidance} and split by measure of \textit{Agility} into those close to industry use (or beyond; \textit{Team Autonomy}) and those with \textit{Reduced Agility} for educational purposes. The \textit{Evaluation} stage deals with patterns related to \textit{Grading}. The position of individual patterns within  categories is a result of their mutual relations and readability of the figure, rather than any further meaning.

\subsection{Source courses}\label{ssec:courses}

The main sources of the patterns themselves and their known uses are the courses we teach at our respective institutions. Among the authors of this paper, we have decades-worth of experience teaching agile practices in particular and software engineering in general. Here we will describe the basic fact of our courses with links to provide the reader with more information in case of interest.

\subsubsection{Software Engineering (ES)}
Software Engineering\footnote{\url{https://sigarra.up.pt/feup/en/UCURR_GERAL.FICHA_UC_VIEW?pv_ocorrencia_id=541882}} (ES) is a course taught at the University of Porto in the spring semester of second year of the Informatics Engineering and Computation Bachelor’s degree programme. Students form teams of 4-6 members working on a software project developing a Flutter app on a topic of their choice (usually within a larger theme chosen by the staff). The project's Scrum-based process is divided into two-week Sprints, plus a longer Sprint 0. 
Weekly lectures are mostly dedicated to teaching agile practices and methods (Unified Process, eXtreme Programming, Scrum, etc.) while also going into topics such as requirements engineering and a bit of UML. Weekly lab classes are dedicated to working on the projects.

\subsubsection{Large Scale Software Development (DS)}
Large Scale Software Development\footnote{\url{https://sigarra.up.pt/feup/en/UCURR_GERAL.FICHA_UC_VIEW?pv_ocorrencia_id=540677}} (DS) is a course at University of Porto for the fall semester of first year of the Master’s degree studies. Each class has a project for an external customer (or with the staff working as the customer), and all teams work on the same implementation (using Large Scale Scrum). The teams consist of 5-6 students using two-week Sprints with an introductory Sprint 0. Weekly lectures are focused on scaled agile practices, while the weekly lab classes are dedicated to project work and team synchronization.

\subsubsection{Advanced Software Engineering (ASWI)}

Advanced Software Engineering\footnote{\url{https://portal.zcu.cz/portal/studium/prohlizeni.html?pc_lang=en}, search Courses with Course abbreviation ASWI} (ASWI) is a course thought at the University of West Bohemia in Pilsen in the second semester of Master's studies. The projects within the course mostly run on a custom process model\footnote{\url{https://shorturl.at/FHCHw} (in Czech)} developed over the year. It is a mash of base Rational Unified Process (RUP)~\cite{kruchten2004rational} structure infused with some agile practices, similar to Disciplined Agile Delivery (DAD)~\cite{ambler2012disciplined}. The project topics are gathered and screened beforehand by the staff from internal and external customers. After assigning the projects, the teams consisting of 3-7 members (with the average of 4) run through the whole project from requirements analysis all the way to release. The iterations range from 1-3 weeks in length (based on the teams' choices).
The lectures cover the entirety of software development process, its practices, foremost process models and methodologies with a focus shifting more towards agile ones through the years. Lab classes exist only in the first weeks to help teams jump onto the process and kick start the projects. Afterwards, there are only mentor consults/evaluation reviews between individual teams and their mentor after each iteration.

\subsubsection{Team Software Project (TSP)}
Team Software Project\footnote{\url{https://portal.zcu.cz/portal/studium/prohlizeni.html?pc_lang=en}, search Courses with Course abbreviation TSP} (TSP) is a year long course (officially split into TSP1 and TSP2, one per semester) taught at the University of West Bohemia in Pilsen. It covers the second and third semesters of Master's studies and its first half runs in parallel to ASWI. Though the course can be taken separately, it most usually serves as an extension of ASWI, both in terms of man-hours and timeline. For students only enrolled into TSP, it only covers basic project milestones and artifacts, leaving most of the process choices to the team. The progress is checked by a process mentor supporting the teams on the practices and processes, as oppose to domain mentors helping with specific technologies. There are no lectures and all the time spent in the course is meant for working on the projects, concluding with presentations at the end of each semester.

\subsubsection{Other courses}
We have gathered further source courses by surveying our colleagues from other institutions and the wider community. The reported comparable courses include:
\begin{itemize}
    \item \textit{\textbf{Team Project (TP)}}\footnote{\url{https://www.cs.ubbcluj.ro/files/curricula/2025/syllabus/IE_sem5_MLE5012_en_dsuciu_2025_9414.pdf}} course taught at Babeș-Bolyai University Cluj-Napoca,
    \item \textit{\textbf{Software Construction (SC)}}\footnote{\url{https://www.fhnw.ch/plattformen/swc/}} course from the University of Applied Sciences and Arts Northwestern Switzerland,
    \item and \textit{\textbf{System Engineering (SE)}} course formerly taught at the Free University of Bozen-Bolzano.
\end{itemize}

\subsection{Pattern template}\label{ssec:template}

The pattern description follows a format with explicitly defined sections. It is a slightly modified version of the template devised by Wellhausen and Fie{\ss}er~\cite{wellhausen2011write} with each pattern having the following:

\begin{itemize}
    \item \textbf{Name} of the pattern
    \item \textbf{Context} -- a situation and circumstances in which the pattern is applied (on top of the ones common to all patterns in the language, see~\ref{ssec:common-context})
    \item \textbf{Problem} -- statement of the problem, which the pattern aims to address
    \item \textbf{Forces} -- the forces causing or complicating the problem that need to be balanced by the solution
    \item \textbf{Solution} -- the resolution to the pattern's problem, balancing the forces (a succinct statement in bold, followed by deeper explanation)
    \item \textbf{Implementation} -- (optional) expands on the specifics or variations of applying the solution
    \item \textbf{Consequences} -- the resulting circumstances of the solution application, either benefits (signalled with a~$+$) and liabilities (signalled with a~$-$)
    \item \textbf{Related Patterns} -- other patterns from the language with identified relation to the current one 
    \item \textbf{Examples / Known Uses} -- particular forms of solutions applied either in source courses (see~\ref{ssec:courses}) or gathered from elsewhere
\end{itemize}

We decided to use a section-based format as it facilitates the reader's comprehension of the pattern contents, facilitating skimming when necessary. This format has also facilitated collaboration between us, the authors, particularly regarding organising our thoughts.

\subsection{Common context}\label{ssec:common-context}

The following common context applies not only to the five patterns described in this paper, but to every pattern in the pattern language:

\begin{quote}
    You are running or planning to establish a university course on software engineering with collaborative student projects and including some aspects or the entire process of agile development. The goal being to provide learning by experience of agile principles and practices as well as general project management skills. No prior knowledge or experience of the students in these domains is assumed\footnote{In cases the course is preceded by others that teach some of these skills, the patterns should be modified accordingly.}.
\end{quote}

\section{Patterns}\label{sec:patterns}

\begin{figure}[htbp]
\centering
\includegraphics[width=0.8\textwidth]{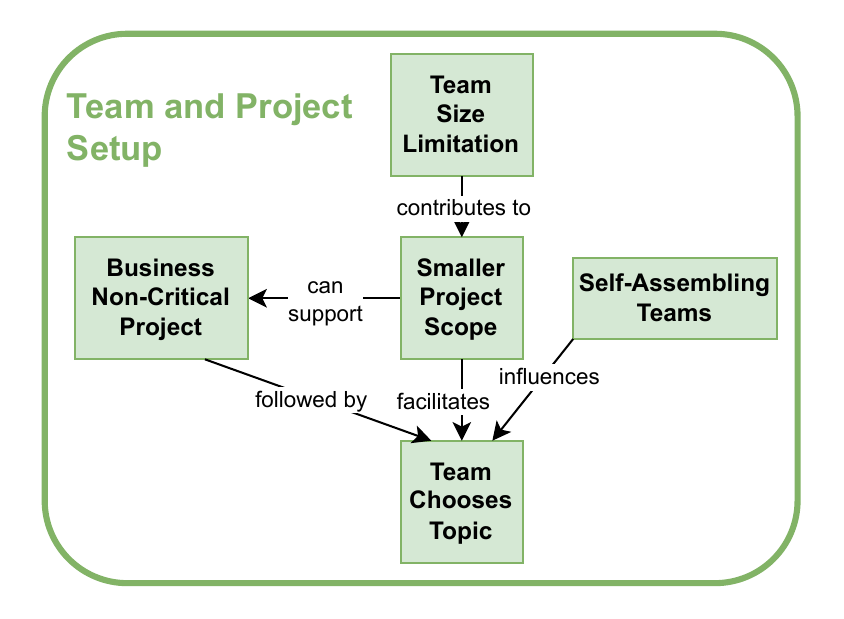}
\caption{Identified relationships between patterns in the Team and Project Setup category.} 
\label{fig-pattern-map}
\end{figure}

Figure~\ref{fig-pattern-map} showcases the relationships between the patterns discussed in this paper, i.e. all from the Team and Project category. 

Table~\ref{tab:paper-patlets} includes the patlets of each pattern discussed in the paper. These are short descriptions for each pattern, outlining the problem they are addressing and showcasing the respective solution.
They are then described in full using the template from~\ref{ssec:template}.

\setlength{\tabcolsep}{5pt}
\begin{table}[htbp]
\caption{Patlets for each pattern in the Team and Project Setup category.}\label{tab:paper-patlets}
\centering
\begin{tabular}{@{}>{\raggedright\arraybackslash}p{0.22\textwidth}p{0.36 \textwidth}p{0.36\textwidth}@{}}
\toprule
\textbf{Pattern}&\textbf{Problem}&\textbf{Solution}\\
\midrule
\textsc{Capping Team Size} & How to best set up the student team size to maximize product and educational efficiency? & Put a reasonable upper limit on team size.\\
    \addlinespace
\textsc{Smaller Project Scope} & How to efficiently teach the practices given the limited capacity of the course? & Assign smaller, worthwhile projects accounting for the educational activities within and the team's capacity.\\
    \addlinespace
\textsc{Business Non-Critical Project} & How not to risk relationship between teachers and customers inherent in students working on core business? & Accept projects that are not on the customer’s critical path or essential for their operations.\\
    \addlinespace
\textsc{Self-Assembling Teams} & How to assemble teams quickly and minimize the initial social and inter-personal hurdles? & Allow teams to self-assemble to create a sense of ownership and freedom of choice.\\
    \addlinespace
\textsc{Team Chooses Topic} & How to pair up the teams with the project topics? & Curate a selection of project topics, but account for the teams' preferences when assigning them.\\
\bottomrule
\end{tabular}
\end{table}

\subsection{Pattern: \textsc{Capping Team Size}}\label{pattern:team-size}

\subsubsection*{Context}
Within the common context (see~\ref{ssec:common-context}), we are setting up the course. The parameters of the projects must be specified, put in the syllabus, communicated to the students enrolling. They are also mutually influential with the rest of the course design.

\subsubsection*{Problem}
How to best set up the student team size to maximize product and educational efficiency? 

\subsubsection*{Forces}
\begin{itemize}
    \item Bigger teams are harder to manage, especially for the inexperienced students, who are either learning or just learned (theoretically) the project management skills.
    \item Agile, specifically, is even harder to grasp, as it is more a mindset than a rigid methodology.
    \item Bigger teams leave open the risk of some members coasting or hiding their minimal contributions.
    \item On the other hand, a team that's too small (and thus, its capacity) reduces the space they have in the project to sufficiently experience and learn project management practices.
    \item Rigidly enforced team size might lead to incomplete teams and lone students given the variable number of students enrolled into the course each year.
    \item Course-specific parameters like class size and learning objectives might play a role.
\end{itemize}

\subsubsection*{Solution}
\textbf{Have an upper and lower limit on (preferred) team size and set the upper boundary low enough.}

Excluding specific exceptions\footnote{Students in distance learning mode are sometimes impossible to align on schedule and group to teams. Foreign exchange students cannot always be grouped neither together nor with local students as they might leave in the middle of the project, for example. Legislation or university policy might demand a student must be allowed to take the course even if (for whatever reason) they cannot be placed in any team.}, the lower limit on team size is apparently two members, as a sole student cannot really be considered a team. Though not ideal, the mechanism of avoiding two-member teams might not be available.

The upper limit is much more within the teacher's discretion. The general advice for team size in industry (cap at 8-10 members~\cite{sutherland2019scrum}), might prove too challenging for novice students in terms of team management practices they are just learning. The ideal seems to be an average team size of around four to six students, but may vary on the course parameters, like class size and learning objectives.

\subsubsection*{Consequences}
\begin{itemize}
    \benefit{Team is still manageable in size for novices while providing ample space to experience and learn project management practices.}
    \benefit{Starting small is in line with agile principles and applying it to team size itself provides further opportunity to showcase this approach to the students.}
    \benefit{The flexible team size and teacher's final say in forming them (see~\textsc{Self-Assembling Teams}, \ref{pattern:self-assembling}) negates the possibility of individual students being left out of the teams.}
    %\benefit{It also allows for better pairing of teams with project topics, which do not necessarily be even in estimated effort.}
    \liability{The smaller team size necessarily limits the scope of the projects the teams can handle, leading to \textsc{Smaller Project Scope}~(\ref{pattern:smaller-scope}).}
    \liability{Enforcing the limits or reforming \textsc{Self-Assembling Teams} \ref{pattern:self-assembling}) to even out the team sizes more or to fit differently estimated project topics might still breed some resentment and frustration among the students right from the start of their projects.}
\end{itemize}

\subsubsection*{Related Patterns}
\begin{itemize}
    \item The pattern contributes to the need for \textsc{Smaller Project Scope}~(\ref{pattern:smaller-scope}) as the team's limited size clearly leads to its reduced capacity to work on the project\footnote{The direction of this relation is education context-specific, as team size is often part of the course design and/or syllabus or a decision by the staff based on available manpower, course objectives and other factors. In industry, in contrast, the project scope would dictate the team size necessary for its completion.}. 
\end{itemize}

\subsubsection*{Examples / Known Uses}
\begin{itemize}
    \item \textit{\textbf{ES}} -- Depending on the number of students in a class group (around 25 students), team size is set to four or five students (in exceptional cases, six students) to have four or five teams of students.
    \item \textit{\textbf{DS}} -- Class groups (of 20-26 students) are divided into four or five teams of 4-6 students of the same size.
    \item \textit{\textbf{ASWI}} -- The team size is set to two to five students, with the most common being four.
    \item \textit{\textbf{TSP}} -- Due to the course running for two semesters, the default team size is set a bit higher to three to seven members, with two-person teams being the (purely circumstantial) exception and the most frequent team size being again four students.
    \item \textit{\textbf{TP}} uses team size of 8-10 people.
    \item \textit{\textbf{SC}} uses teams of maximum 3 students.
\end{itemize}
\subsection{Pattern: \textsc{Smaller Project Scope}}\label{pattern:smaller-scope}

\subsubsection*{Context}
On top of the common context (see~\ref{ssec:common-context}), after deciding on the \textsc{Capping Team Size}~(\ref{pattern:team-size}), another course aspect to consider is the scope of the projects the student teams will be working on.

\subsubsection*{Problem}
% Whether the project happens simultaneously with theoretical education or after it, the time in the course is limited and learning is focused on the practices and the process employed in the project, not strictly on the end product. In addition to lectures (if included in the course), learning through experience can be a worthwhile approach to take, but students may focus their learning on the less focal topic (i.e., product rather than process) during their project.
How to efficiently provide students with practical education on the process and practices while creating a worthwhile product given the limited time the students have as dictated among others by the credit volume of the course?

\subsubsection*{Forces}
\begin{itemize}
    \item The students do not work full time on the project, juggling it with their responsibilities from other parts of the course (lectures, exams; plus the overall spent time limitation on the course per the number of credits gained from passing), other courses, personal lives and, potentially, part-time jobs. Depending on how many credits the course is worth they may have more time, but there is no exclusivity on the students’ end.
    \item Agile processes combine several moving parts, including ceremonies, practices, and principles, to name a few. Hands-on experience with these helps students see how everything fits, leading to better comprehension and a sense of being more than just theoretical stuff they need to recite for the exam.
    \item Students should have something to work towards, and a project with a clear goal is a good way to keep them engaged. Even better if they can see a real-world benefit to a customer who validates their effort and results, rather than "yet another" artificial (purely) educational assignment which will only end up in a drawer forever.
    \item Students may be hardwired from other courses that all that matters for the project is the quality of the end result, automatically ignoring the "whys and hows" to reach it reliably.
\end{itemize}

\subsubsection*{Solution}
\textbf{Assign smaller projects to students that can be of value while accounting for the team's size, the expected engagement from the course, and the educational activities within the project.}

When we, as educators, want to teach students about the mechanisms behind the processes they are working on, we should focus their attention at the mechanisms themselves. However, it is easy for students to not see the forest for the trees depending on their previous experiences in other courses, where the project's end result was all that mattered.

As such, in order to shift the students attention, educators can propose topics that are \emph{smaller} in scope but that still make for a coherent and well-thought out project.

With this, students should not be working fully at capacity, giving them room for agile ceremonies, other tasks (such as the ones that may be present for evaluation purposes but that do not advance development), and also giving them slack time. In a course where each student should have 80 hours of development time, they should be working on a 60-hour project.

This can be achieved, in the case of an externally-proposed topic, by giving students room to negotiate scaling down the scope of the topic. On the other hand, if the students are proposing their own topic, they can start small, having room to grow with the guidance of their professor.

\subsubsection*{Consequences}

\begin{itemize}
    \benefit{Having projects that are smaller in scope helps students to maintain a sustainable pace, helping them remain aligned with agile principles. The sustainable pace assists in students not scrambling to get things done at the last minute, which together with a reduced workload gives students the room to be intentional about their actions and really take in the reasons behind what they are doing in the course of their projects.}
    \benefit{Smaller projects are also more easily and accurately grasped mentally. This can spark creativity, when students are given a choice to come up with their own project topics, or help assess the topics assigned or up for selection.}
    \liability{A reduced project scope can backfire and steer students the wrong way, downplaying what they need to work on for the project and resulting in poorer time management, which in turn leads the team to work at an unsustainable pace.}
\end{itemize}

\subsubsection*{Related Patterns}

\begin{itemize}
    \item The need for smaller project scope is partially driven by the \textsc{Capping Team Size}~(\ref{pattern:team-size}) as the aggregate team capacity puts a limitation on the scope from the start\footnote{The note on this relation from \textsc{Capping Team Size}~(\ref{pattern:team-size}) applies here.}.
    %\item Having a smaller scope can benefit instances of a \textsc{Process-Focused Evaluation}, as it helps drive home to students the notion of focusing on the process and not on the final result.
    \item These smaller projects can also facilitate \textsc{Team Chooses Topic}~(\ref{pattern:team-chooses-topic}). In case of a self-suggested topic by the team, it can spring forth creativity -- creating "something that does one thing very well". In case of picking from pre-prepared topic selection, it can at least help students in comprehending and assessing the complexity and overall effort needed for the project. In either case, it enables them to make a better informed decision.
    \item The smaller scope also supports the \textsc{Business Non-Critical Project}~(\ref{pattern:business-non-critical-project}) topic, as the customers rarely have proposals that are small enough for the courses, yet very important or business critical.
\end{itemize}

\subsubsection*{Examples / Known Uses}

\begin{itemize}
    \item \textit{\textbf{ES}} -- Teams are encouraged to keep their project topics simple with an easy-to-implement minimum viable product, with the option to grow in scope as the semester progresses and the team implements their initial view of the project. Students are expected to work around 60-75 man-hours per team member during the semester, with the project's scope being adjusted to the team's size.
    \item \textit{\textbf{ASWI, TSP}} -- Customers are informed by the course staff prior to topic submissions on the estimated scope of the projects, taking into account the administrative overhead the students have to apply in the projects~(\textsc{Higher Ceremony}, see Appendix \ref{ap:patlets}). They are also encouraged to submit topics that can be easily scaled up or down based on the specific project context (i.e., the varying number of team members). The topics are subsequently screened by the course staff in this regard to the best of their ability. The mentors are also ready to assist the teams to resolve a situation where the fact that the topic is significantly more/less complex than originally estimated is discovered during the project's execution. In terms of concrete targeted scope in man-hours (mh), the courses differ:
    \begin{itemize}
        \item ASWI -- 60-80mh per team member,
        \item TSP -- 80-100mh per student, per semester,
    \end{itemize}
    Thus, if we imagine a case of team of four members who are all enrolled in both subjects, the total expected scope is 880-1120mh. On average, around a third of the time, however, is spent on ceremonies, overhead, and other activities needed more for learning than the product itself. 
    \item \textit{\textbf{TP}} uses teams of 8-10 students with the expected 50 hours spent per student per semester.
    \item In \textit{\textbf{SC}} teams of 3 students work on a small, approx. 4K LOC, application.
\end{itemize}
\subsection{Pattern: \textsc{Business Non-Critical Project}}\label{pattern:business-non-critical-project}

\subsubsection*{Context}
In addition to the common context (see~\ref{ssec:common-context}), through \textsc{Capping Team Size}~(\ref{pattern:team-size}) and \textsc{Smaller Project Scope}~(\ref{pattern:smaller-scope}), we now need a mechanism for coming up with the actual project topics to assign to teams.

\subsubsection*{Problem}
How to minimize the risks inherent in students working on parts of the customer's core business?

\subsubsection*{Forces}
\begin{itemize}
    \item Students need a topic to work on, and industry-oriented ones have their unique interesting factors, such as opportunities to interact with external customers and the potential to use new or different technologies.
    \item Companies and other organisations may be willing to collaborate with student projects seeking free manpower or to start relationships that may lead to future hiring. However, this goal often takes second place when faced with the daily work and demands of the business, sometimes resulting in limited availability.
    \item The students, as well as the course staff, cannot guarantee the extent and quality of the products the teams deliver as the product is at best a co-equal, often even secondary, goal next to practical education in processes and practices.
    \item In case of IT customer, their product release schedules may not align with course's needs.
    \item Customers may further hesitate to take part in student projects due to risk of inexperienced students with no legal responsibility unintentionally interfering with business critical assets, leaking trade secrets and other confidential data.
    \item Potential failure or overall bad experience with the projects may jeopardize the working relationship between the teaching staff and the customers, either internal (colleagues) or external (industry partners).
    \item The risk of failure in something crucial increases stress in all parties involved, but especially the students due to their lack of experience. 
\end{itemize}

\subsubsection*{Solution}
\textbf{Accept projects that are not on the customer’s critical path or essential for their operations.}

Rather then customer's business products themselves, these may include nice-to-have utilities, research tools, quality of life software for non-IT customers, etc.

When outside organisations decide to assume the customer role, they are giving students the opportunity to get a taste of "the real world", with them being able to form connections with practitioners and learn more about how different organisations operate, all in addition to practising agile methods and learning through iterative delivery.

Giving customers an opening to propose project topics that fall outside of their main products and features can give them this opportunity to build bridges with the university and the people who will be entering the workforce in the future, while minimising the inherent risk that comes with student projects --- where the quality of the end result can vary to the point that there may not be much there at all.

%\subsubsection*{Implementation}

\subsubsection*{Consequences}

\begin{itemize}
    \benefit{The non-criticality of projects enables students and educators to be connected to outside organisations, while reducing any risks that come from the variable quality of student projects. These opportunities can help kickstart the students' future careers, as they come into the workforce with connections and a world-view that is closer to the practitioners' reality.}
    \benefit{If customers' expectation is lowered and not critically dependent on the projects' results, the room for disappointment shirks, yet there is still space to be pleasantly surprised.}
    \benefit{Non-critical topics reduce stress from consequences of a potential failure.}
    \liability{Companies may not be interested in proposing topics that are not useful for them; as they would be taking on the role of a customer, the availability required of them to be in meetings and work with the students may not be worth it for them.}
\end{itemize}

\subsubsection*{Related Patterns}
\begin{itemize}
    \item Having companies propose project topics that are not their main, critical path can actually produce ideas with a \textsc{Smaller Project Scope}~(\ref{pattern:smaller-scope}).
    \item These customer-proposed topics can be a source of curated projects students can choose from in \textsc{Team Chooses Topic}~(\ref{pattern:team-chooses-topic}).
    \item Due to the non-critical nature of the projects and their time constraints, the students might not have the access to the end users of the developed software (i.e., the customer's customers), or at least their fully representative sample, or the time to gain it. This may contribute to the deployment of the \textsc{Shallow Feedback Loops} (see Appendix~\ref{ap:patlets}).
\end{itemize}

\subsubsection*{Examples / Known Uses}
\begin{itemize}
    \item \textit{\textbf{DS, ASWI, TSP}} -- When gathering topics for projects prior to their publishing to the student teams to for expressing interest, the customers are advised by the course staff to not submit any business critical topics, as the results as their quality cannot be properly guaranteed. This results in mostly nice-to-have applications, extensions, integrations, etc. being proposed by the external customers and supporting software for research projects larger in scope than individual thesis topics being submitted by internal customers.   
    \item In \textit{\textbf{TP}}, the project topics are often proposed by industry specialists but have limited significance and serve primarily for educational purposes.
\end{itemize}
\subsection{Pattern: \textsc{Self-Assembling Teams}}\label{pattern:self-assembling}

\subsubsection*{Context}
Within the common context (see~\ref{ssec:common-context}), new students enrol into our course. The semester has either just begun or is close to it. The actual work on the projects should start in the near future.

\subsubsection*{Problem}
How to assemble the teams from the individual students quickly and in a way that minimizes the initial social and inter-personal hurdles?

\subsubsection*{Forces}
\begin{itemize}
    \item Teams need to be formed relatively quickly for the work on the projects to start.
    \item Being forced to cooperate with newly acquainted or unfitting team members (on personal or professional level) can lead to resentment, tension and frustration in the team.
    \item Some, but not all, students formed interpersonal relationships in their previous studies or personal lives and have a pallet of (soft and hard) skills each that can be complementary or contrary to each other, none of which the course staff is necessarily aware of.
    \item Trying to assemble the "perfect" teams puts a lot of pressure and effort on the course staff in the setup stage, especially if the number of students is high. Furthermore, the underlying data for such an effort might be missing or unreliable.
    \item The teams need to meet certain general (e.g., \textsc{Capping Team Size}, see~\ref{pattern:team-size}) or course-specific criteria.
    \item A completely hands-off approach to team formation can lead to breaking the aforementioned criteria, delays, or leave some students out in the cold.
\end{itemize}

\subsubsection*{Solution}
\textbf{Allow and encourage the students to form the teams on their own at the very beginning of the course or (if possible) even before.}

After all, self-organizing teams (in terms of internal roles and structure) are one of the staples of Agile and boosting this up to the level of team assembly itself helps strengthens this principle in the minds of the students.

Make them aware of any criteria that need to be met and give them a near enough deadline to "register" their team composition and put a approval/vetting process in place to review the teams formed. After the deadline has passed, resolve any cases where full teams were not formed -- such as incomplete teams or isolated students without a team -- using accessible knowledge and your best judgment. 

\subsubsection*{Consequences}
\begin{itemize}
    \benefit{Teams form quicker when students leverage their interpersonal relationships and prior experience with each other, compared to if the staff needed to learn about them first.}
    \benefit{Joining a team voluntarily, as opposed to getting stuck in one by a decree from authority, is one less frustration students need to deal with with navigating the already tricky realm of team dynamics and collaboration.}
    \benefit{Self-assembled teams can feel higher measure of ownership and autonomy of their team and the project as a whole.}
    \benefit{Lower time and effort investment on the staff side, and requirements for prior knowledge on individual students skills, experience and personalities.}
    \benefit{Due to the vetting process, any and all criteria for the teams are still met, or at the very least, their exceptions are monitored and regulated by the course staff.}
    \benefit{The chance of a lone student without a team needing to work on a solo project is dramatically lowered.}
    \liability{Teams self-assembled based on pre-existing friendships while ignoring skill sets may cause issues.}
    \liability{Some resentment can still arise in those teams forced to accept lone students, or those formed entirely from such individuals.}
    \liability{Even voluntarily formed teams may face issues in team dynamics and cooperation, which still needs mediation and other actions from the course staff.}
\end{itemize}

\subsubsection*{Related Patterns}
\begin{itemize}
    \item The pattern influences the \textsc{Team Chooses Topic}~(\ref{pattern:team-chooses-topic}) of the project topic (or at least the expressed preferences) as all team members need to sign-off on, or at least concede to, the collective decision.
    \item After assembling into teams, the students should also agree on their individual roles and take partial ownership of the details of their process, which they share with the mentor (staff member), who sets the more coarse outline of the process. Thus, taking part in \textsc{Dual Process Ownership} (see Appendix~\ref{ap:patlets}).
\end{itemize}

\subsubsection*{Examples / Known Uses}
\begin{itemize}
    \item \textit{\textbf{ES, DS}} -- Students in the same lab class self-assemble in teams depending on the team size defined by the teachers during the first lab class. When teams deviate from the expected size (either when a team has too many students or there are students without a team), the teacher intervenes and facilitates team assembly with the students.
    \item \textit{\textbf{ASWI}} -- Students are told to form the teams of the expected size on their own at the very beginning of the term and given at most two weeks to do so. As part of their registration they also assign the team lead (or at least a contact person), express their preferences for project topics (see~\textsc{Team Chooses Topic}, \ref{pattern:team-chooses-topic}) and provide their team profile in regards experience with different technologies. The course staff vet the team registrations and make any changes as necessary. Any students unassigned to any team by the deadline (if any) are either assigned to as existing team or grouped together to new teams. Either is done based on the target team size, technical experience compatibility, preferred topics and other criteria.
    \item \textit{\textbf{TSP}} -- The process is broadly similar to the one used in ASWI. Only it starts several months before the term starts and it is supported by a custom web application for team registrations and searching for teammates. If the team is at least partially enrolled in both courses, TSP version is followed.
    \item In \textit{\textbf{TP}} project teams used to be formed randomly or based on various other criteria, but participants' motivation was negatively affected. Now, they use self-assembling teams following exactly this pattern.
    \item \textit{\textbf{SC}} uses an online sign-up sheet for students to form teams.
    \item In \textit{\textbf{SE}}, students are free to form the teams on their own. 
\end{itemize}
\subsection{Pattern: \textsc{Team Chooses Topic}}\label{pattern:team-chooses-topic}

\subsubsection*{Context}
We are in the common context (see~\ref{ssec:common-context}). We have the students grouped into teams. We also have the curated list of feasible topics satisfying all the criteria for the projects. In other words, \textsc{Smaller Project Scope}~(\ref{pattern:smaller-scope}), \textsc{Business Non-Critical Project}~(\ref{pattern:business-non-critical-project}) and \textsc{Self-Assembling Teams}~(\ref{pattern:self-assembling}), or their valid equivalents, have already been applied.

\subsubsection*{Problem}
The best situation is a team working on a project their are enthusiastic about and qualified for. How to find the best fit when assigning the assembled teams to the gathered project topics?

\subsubsection*{Forces}
\begin{itemize}
    \item Having all students work on topics that are consistent in terms of difficulty helps with evaluation, enabling comparable evaluation criteria.
    \item The topics the students work on should be adequate for the course, in terms of scope, complexity, number of features, and tech stack.
    \item Feeling motivated helps students bring out their best work, but extrinsic motivation (such as that fuelled by a good grade) is not enough; intrinsic motivation~\cite{deciSelfDeterminationTheory2012} (related to personal development and attachment to the work in progress) is paramount in a software project~\cite{beecham2008MotivationSoftwareEngineering}.
    \item The staff is not equipped to judge the teams' preferences, confidence, ambitions and experiences with different technologies without adequate information.
\end{itemize}

\subsubsection*{Solution}

\textbf{Curate the selection of project topics, but take into account the preferences of team members when assigning them.}

The project topics available to students should be managed by the professors, enabling different topics to be consistent in terms of scope, technology, and complexity.

%The collection of topics may come up through various sources, which may depend on the course's learning objectives and approaches. A course that aims to give students experience with working with industry practitioners may have topics proposed by external customers, while a course that focuses on providing students with core competencies with project management may use internally-proposed project topics. 

%The history team members have with each other (if any) can influence how the team chooses their topic. For instance, if students had the opportunity to self-organise in teams and chose colleagues they are already friendly with or had worked previously, they will have a stronger familiarity and team identity compared to other teams where the students are not really familiar with each other. 
There might be a discrepancy in pace of teams forming due to pre-existing relationships between students, or lack there of. To give all teams a level play field, with the opportunity for them to go through the team forming process~\cite{tuckmanStagesSmallGroupDevelopment1977}, the task of choosing a project topic ought to not be immediate. Teams should have time (depending on the course calendar, two or three weeks at the course's beginning or even before its start) to assess their options and make the choice they feel is best.

Regardless of the origin of the project topics, students should have a say on which topic they will be working on during the semester.

\subsubsection*{Implementation}

The following are alternative ways of applying the general solution in a still generic-enough way:
\begin{itemize}
    \item Let students come up with their own project topics. With adequate guidance on the professor's part, students can come up with a topic they are invested in and that also meets the course's learning objectives.
    \item Let students bring topics from their own contacts gained through previous studies, participation in research, personal relationships with people in the industry, or jobs they might already have.
    \item Prepare a pool of topics and let students pick from it. These topics can be proposed internally, by the professors, or externally, by outside organisations and customers. This can give the professors more control over the consistency between different topics, while giving students agency on their work. 
\end{itemize}
Any mixture of all of the above is obviously also possible, e.g., have a prepared pool of topics as a default, but allow (or even encourage) students to bring or come up with their own proposals.

\subsubsection*{Consequences}

\begin{itemize}
    \benefit{Being able to have a say in what they will work on helps with students' sense of ownership, commitment and buy-in.}
    \liability{Giving the teams a choice about their topics requires some care. In an instance where students propose their topics, they may end up with contrasts between teams, with some topics being more complex or wider in scope than others. This can make it harder to evaluate the students' work due to the different playing fields.}
    \liability{If the description of a topic is too surface-level, short or vague, it can lead to mismatch between expectations and reality in either direction on the students' or teachers' part. For example, the topic can appear easy to students and turn up to be overly challenging for their experience level. Or the topic (no matter its source) can be assessed as complex enough by the staff when its significant portion can be easily addressed by a free, third-party solution, thus drastically cutting the effort needed for the overall product.}
    %\liability{The course's (and, consequently, the project's) time scale may be small to the point where students do not have all the time they should have to go through the team forming process and choose their topic with time and care. In these cases, professors can either assign students to teams using their own discretion (eliminating disparities in team familiarities that can show up with self-organised teams) or try to include team-building activities in the program before students move on to choosing their topics.}
\end{itemize}

\subsubsection*{Related Patterns}
\begin{itemize}
    \item Giving teams a choice over the topics can be made easier by ensuring a \textsc{Smaller Project Scope}~(\ref{pattern:smaller-scope}), as students can better understand the point of the project and what they will be doing on them, as well as compare between different topics, if these are small.
    \item The history team members have with each other (if any) can influence how the team chooses their topic. For instance, if students had the opportunity to form \textsc{Self-Assembling Teams}~(\ref{pattern:self-assembling}) and chose colleagues they are already friendly with or had worked with previously, they will have a stronger familiarity and team identity compared to other teams where the students are not really familiar with each other. That can lead to a more confident and appropriate topic choice or an idea for their own project.
    \item Apart from the above, the pattern obviously follows the collection of \textsc{Business Non-Critical Project}~(\ref{pattern:business-non-critical-project}) topics. 
\end{itemize}

\subsubsection*{Examples / Known Uses}

\begin{itemize}
    \item \textit{\textbf{ES}} -- The course’s teachers select a common, broad theme for the project topics. In previous years, student apps have been related to student life, and more recently to the UN’s Sustainable Development Goals\footnote{\url{https://sdgs.un.org/goals}}. Students then choose a specific topic within these broader themes and receive feedback from their teachers throughout the process.
    \item \textit{\textbf{ASWI, TSP}} -- A "call for topics" is sent out to both the industrial partners and internal organizational units of the university by the course staff well in advance (i.e., before the semester in which the course is taught starts). The gathered topics are screened for complexity, estimated scope (\textsc{Smaller Project Scope}, see~\ref{pattern:smaller-scope}), educational value, technology, business criticality (\textsc{Business Non-Critical Project}, see~\ref{pattern:business-non-critical-project}), etc. The students are also allowed to bring their own topic proposals, but a customer from outside the team needs to be defined. The resulting set is published to the students and the teams can express their preferences. The staff then assigns the projects mainly on these, outside of collisions in preferences or a stark mismatch of the topic and team's profile (technology experience, etc.). 
    \item In \textit{\textbf{TP}}, the project teams pick a topic proposed by industry specialists or they define the project topic themselves, while project's difficulty is assessed by and instructor to remain consistent across teams. Project topics are not imposed, as this would decrease motivation for implementing the project.
    \item In \textit{\textbf{SE}}, students were free to choose their own topic. 
\end{itemize}

\section{Conclusions and future work}\label{sec:conclusions}

In this paper we presented the beginnings of a pattern language that focuses on ASD practices in higher education context, with five patterns for team and project setup forming an initial subset: \textsc{Capping Team Size}~(\ref{pattern:team-size}), which advised composing even smaller teams than is usual in industry; \textsc{Smaller Project Scope}~(\ref{pattern:smaller-scope}), which focused on how smaller projects can help students learning the different moving parts in the process; \textsc{Business Non-Critical Project}~(\ref{pattern:business-non-critical-project}), which discussed how external customers can reduce the risks that come with student projects while maintaining the benefits students get when working with real-world customers; \textsc{Self-Assembling Teams}~(\ref{pattern:self-assembling}), which endorsed students' freedom to form the teams to boost cooperation; and \textsc{Team Chooses Topic}~(\ref{pattern:team-chooses-topic}), which dived into the benefits of letting students be involved in choosing the topics they work on. The rest of the patterns proposed for the language are captured in patlet form in Appendix~\ref{ap:patlets}.

We recognize that some, if not all, of the patterns may be applicable in contexts outside of ours, e.g., non-agile courses, projects shorter than a semester, even industry in some cases. However, given we observed these patterns in a specific context (see~\ref{ssec:courses} and \ref{ssec:common-context}), we make no definite proclamations regarding their wider applicability.

We further argue that the application of the solutions is more beneficial to agile courses than others. As the agile mindset is more challenging to teach and/or learn than a rigidly prescribed methodology, the students benefit from learning on smaller teams to manage, necessarily related to smaller project scopes, less critical assignments with less stress of failure and the option to choose their teammates as well as the topic they feel comfortable with.

In the future we plan to expand this budding pattern language with more patterns, fully describing the rest of our mined set (see Appendix~\ref{ap:patlets}) and potentially capturing more, diving into the relationships between them, exploring the different practices employed by educators, and learning more about how students can learn about ASD in a university environment.

\section*{Acknowledgements}

The authors would like to thank the shepherd of this paper, Tineke Jacobs, as well as Workshop Group G at EuroPLoP '25, for their work, valuable feedback and overall help in improving the patterns and their presentation.

Further thanks to Dan-Mircea Suciu, Eduardo Guerra and Martin Kropp for providing additional known uses for the individual patterns.

\appendix
\section{Pattern language patlets}\label{ap:patlets}

This appendix includes the patlets for all patterns for the language, grouped by their categories.

\subsection{Category: Team and Project Setup}
\subsubsection{\textsc{Capping Team Size}}
\begin{description}
    \item[Problem:] How to best set up the student team size to maximize product and educational efficiency?
    \item[Solution:] Put a reasonable upper limit on team size.
\end{description} 

\subsubsection{\textsc{Smaller Project Scope}}
\begin{description}
    \item[Problem:] How to efficiently teach the practices given the limited capacity of the course?
    \item[Solution:] Assign smaller, worthwhile projects accounting for the educational activities within and the team's capacity.
\end{description}     
    
\subsubsection{\textsc{Business Non-Critical Project}}
\begin{description}
    \item[Problem:] How not to risk relationship between teachers and customers inherent in students working on core business?
    \item[Solution:] Accept projects that are not on the customer’s critical path or essential for their operations. 
\end{description}          

\subsubsection{\textsc{Self-Assembling Teams}}
\begin{description}
    \item[Problem:] How to assemble teams quickly and minimize the initial social and inter-personal hurdles?
    \item[Solution:] Allow teams to self-assemble to create a sense of ownership and freedom of choice.
\end{description}

\subsubsection{\textsc{Team Chooses Topic}}
\begin{description}
    \item[Problem:] How to pair up the teams with the project topics?
    \item[Solution:] Curate a selection of project topics, but account for the teams' preferences when assigning them.
\end{description}

\subsection{Category: Guidance}
\subsubsection{\textsc{The Mentor-Assessor Role}}
\begin{description}
    \item[Problem:] How to perform both guiding the team and grading their performance, which both require deep insight into inner team dynamics, practice applications and communication (internally and externally), while saving on manpower and scheduling difficulties?
    \item[Solution:] Have a single person embody both the mentor and the assessor role. 
\end{description} 

\subsubsection{\textsc{Big Brother}}
\begin{description}
    \item[Problem:] How to grant the mentor/assessor access to detailed information of inner workings of the team, to get the full picture for evaluation and insight for mentoring?
    \item[Solution:] Unless this conflicts with information sensitivity on customer’s side make even internal team notes/communications freely accessible to the mentor/assessor.
\end{description}

\subsubsection{\textsc{Early Meetings Crutch}}
\begin{description}
    \item[Problem:] How to facilitate hands on learning for good practices that are hard to explain otherwise?
    \item[Solution:] In early team meetings, let the mentor be a fly on the wall and subsequently give team a constructive feedback on their conduct, structure and communication skills. 
\end{description}

\subsubsection{\textsc{Check Customer's Privilege}}
\begin{description}
    \item[Problem:] How to mitigate the customer's (especially IT-industrial ones) tendency to “take over” the project and play roles that the team should learn by doing, e.g., requirements analyst, architect and project manager?
    \item[Solution:] Inform and enforce the role the customer should stick to, including "playing dumb", and help the team in situation when customer oversteps.
\end{description} 

\subsection{Category: Process Tailoring -- Reduced Agility}
\subsubsection{\textsc{Higher Ceremony}}
\begin{description}
    \item[Problem:] How can we reliably verify the students have tried out and learned all the standard practices, even those that would not be call for producing hard evidence (paper trail) in industry due to the projects' smaller scale? 
    \item[Solution:] Explain this dilemma and the educational reasons for the level of ceremony to the students openly and clearly. Set the process to necessitate the taught practices and tools to a sufficient but not extremely overwhelming degree.
\end{description}

\subsubsection{\textsc{Disproportional Initial Stage}}
\begin{description}
    \item[Problem:] How to ease the burden on new student teams  who struggle to balance preparation and execution when thrown directly into iterative development cycles?
    \item[Solution:] Allocate enough time at the beginning of the project to allow teams to establish a shared understanding, set up tools, and practice agile fundamentals without the pressure of immediate delivery. 
\end{description}

\subsubsection{\textsc{Full Product Backlog from Day One}}
\begin{description}
    \item[Problem:] How to help the teams who may be lost in terms of what to work on first at the point they do their first team planning?
    \item[Solution:] Push and help the teams to define a set of features/user stories to work on throughout the project, having a general idea of what will be worked on at which time.
\end{description}

\subsubsection{\textsc{Non-Daily Scrum}}
\begin{description}
    \item[Problem:] How to resolve the need to showcase the usefulness of stand-ups with the potential impracticability of their daily performance due to time constraints or conflicting schedules, which could lead to skipped or poorly attended meetings that undermine team alignment?
    \item[Solution:] Introduce a cadence for stand-ups that matches the team's capacity and work rhythm, such as (bi-)weekly, while maintaining the focus on frequent updates and collaboration. 
\end{description} 

\subsection{Category: Process Tailoring -- Team Autonomy}
\subsubsection{\textsc{Dual Process Ownership}}
\begin{description}
    \item[Problem:] How to help inexperienced students with potentially the most challenging and simultaneously most educationally potent role of Scrum Master (or equivalent)?
    \item[Solution:] Use a mentor, an experienced staff member, as a crutch to help oversee, tailor and adhere to the team’s process. Simultaneously, have an internal process owner (an individual or the whole team, their choice), as the mentor cannot be present all the time.  
\end{description} 

\subsubsection{\textsc{Self-Defined Policies}}
\begin{description}
    \item[Problem:] How to device detailed team policies (on code, notations, roles, tools, planning strategies, etc.) without rigidly imposing them, risking dismissal, resentment and blockers to learning through experimentation?
    \item[Solution:] Define just as much of the process as necessary (for comparisons and manageability on the course level) and give students free reign over the rest. Also, be prepared to bend or change even the process structure you do have if they come with a justified case for it and explicitly encourage such ideas.
\end{description} 

\subsubsection{\textsc{Strictly Time-Boxed Team Ceremonies}}
\begin{description}
    \item[Problem:] How to help focus meetings, so people, especially inexperienced students, do not wonder off from the topic at hand wasting everyone’s time?
    \item[Solution:] Leverage limited class time to have hard time constraints on routine meetings with specific purpose (e.g. Sprint Review and Sprint Planning). 
\end{description}

\subsubsection{\textsc{Shallow Feedback Loops}}
\begin{description}
    \item[Problem:] How to ensure timely feedback on the product increments that the teams provide when a non-business critical nature of the project and unavailability of end users may cause delays? 
    \item[Solution:] Ask customers to act as proxy users who provide feedback on their behalf.
\end{description}

\subsubsection{\textsc{Cross-Pollination}}
\begin{description}
    \item[Problem:] How to provide students with the advice/experience they might be more receptive to than the one from a perceived prescriptive authority (a teacher)?
    \item[Solution:] Allow teams to learn and draw inspiration from each other through opening internal team data and mentor evaluations to the whole class, casual contact between members of different teams, and/or organized experience exchange events.
\end{description} 

\subsection{Category: Grading}
\subsubsection{\textsc{Process-Focused Evaluation}}
\begin{description}
    \item[Problem:] How to properly assess the quality of work and learning outcomes in a course of this nature?
    \item[Solution:] Base the larger part of students’ evaluation on the handling the process and practices, attitude and problem solving during project management, not the product and its quality. 
\end{description} 
    
\subsubsection{\textsc{Not All Iterations Are Equal}}
\begin{description}
    \item[Problem:] How to account for the learning curve at the start of the project and the necessary ceremonies that might exist at the end of it (due to the educational context) in the evaluation process?
    \item[Solution:] Put less weight in evaluation on the temporal extremes of the project takong their specific context into account.
\end{description}

%
% ---- Bibliography ----
%
% BibTeX users should specify bibliography style 'splncs04'.
% References will then be sorted and formatted in the correct style.
%
 \bibliographystyle{splncs04}
 \bibliography{references}
\end{document}